\documentclass{article}
\usepackage{amsfonts}
\usepackage{amsmath}
\usepackage{graphicx}
\usepackage{cite}
\usepackage{hyperref}

\begin{document}

\title{Asymmetric Duffing oscillator: 
metamorphoses of $1:2$ resonance and its interaction with the primary resonance
}
\author{Jan Kyzio{\l }, Andrzej Okni\'{n}ski \\
Politechnika \'{S}wi\c{e}tokrzyska, Al. 1000-lecia PP 7,\\
25-314 Kielce, Poland}
\maketitle

\begin{abstract}
We investigate the $1: 2$ resonance in the periodically forced asymmetric Duffing oscillator due to the 
period-doubling of the primary $1: 1$ resonance or forming independently, coexisting with the primary resonance. 
We compute the steady-state asymptotic solution -- the amplitude-frequency implicit function.
 Working in the differential properties of implicit functions framework, we describe complicated metamorphoses 
of the $1:2$ resonance and its interaction with the primary resonance.
\end{abstract}

\section{Introduction and motivation}
\label{introduction}

A period-doubling cascade of bifurcations is a typical route to chaos in
nonlinear dynamical systems. A generic example is the asymmetric Duffing
oscillator governed by the non-dimensional equation
\begin{equation}
\ddot{y}+2\zeta \dot{y}+\gamma y^{3}=F_{0}+F\cos \left( \Omega t\right) ,
\label{Asym-Duffing}
\end{equation}%
which has a single equilibrium and a corresponding one-well potential \cite%
{Kovacic2011}, where $\zeta $, $\gamma $, $F_{0}$, $F$ are parameters and $%
\Omega $ is the angular frequency of the periodic force.

 Szempli\'{n}ska-Stupnicka elucidated the period-doubling scenario in the dynamical system 
(\ref{Asym-Duffing}) in a series of far-reaching papers \cite{Szemplinska1986,Szemplinska1987,Szemplinska1988}; 
see also \cite{Kovacic2011} for a review and further results.

The main idea introduced in \cite{Szemplinska1986} consists of perturbing
the main steady-state (approximate) asymptotic solution of Eq.(\ref%
{Asym-Duffing})%
\begin{equation}
y_{0}\left( t\right) =A_{0}+A_{1}\cos \left( \Omega t+\theta \right) ,
\label{AD-solution}
\end{equation}
as
\begin{equation}
y\left( t\right) =y_{0}\left( t\right) +B\cos \left( \tfrac{1}{2}\Omega
t+\varphi \right) ,  \label{y}
\end{equation}%
substituting $y\left( t\right) $ into Eq.(\ref{Asym-Duffing}) and
considering the condition $B\neq 0$. In papers \cite%
{Kovacic2011,Szemplinska1986,Szemplinska1987,Szemplinska1988}, 
the authors found several conditions guaranteeing the formation and stability of 
solution (\ref{y}) and used them to study the period-doubling phenomenon.

In our recent work, we studied the period-doubling scenariousing the period-doubling 
condition determined in \cite%
{Kovacic2011,Szemplinska1986,Szemplinska1987,Szemplinska1988} as an
implicit function.
More precisely,
using the formalism of differential properties of implicit functions \cite%
{Fikhtengolts2014,Kyziol2022}, we derived analytic formulas for the birth of
period-doubled solutions \cite{Kyziol2023b}.

The motivation of this work stems from two observations: (i) in some cases, $1:2$ resonance, 
coexisting with $1:1$ resonance, is not created via the period doubling of $1:1$ resonance, 
(ii) $1:2$ resonance depends 
on the parameters in a more complicated way than the primary resonance. 

Thus, the aim of the present work is to study metamorphoses of $1:2$ resonance coexisting with 
the primary $1:1$ resonance.

The paper is structured as follows: Section \ref{solution} describes amplitude-frequency curve for 
$1:2$ resonance for Eq. (\ref{Asym-Duffing}). In Section \ref{singular}, equations to 
compute singular points and vertical tangencies are derived. In Section \ref{numerical}, we present an example 
of metamorphoses of $1:2$ resonance, based on computed singular points. 
In Section \ref{test} we provide an example of similar metamorphoses for a different dynamical system, 
suggesting a greater generality of our results.  
We summarize our results in the last section.

\section{The $1:2$ resonance: steady-state solution}
\label{solution}
Since the $1:2$ resonance can coexist, without contact, with the primary $1:1$ resonance 
we assume the following steady-state solution of Eq.(\ref
{Asym-Duffing})

\begin{equation}
y\left( t\right) =B_{0}+B\cos \left( \tfrac{1}{2}\Omega t+\varphi \right) ,
\label{Solution 1/2}
\end{equation}%
which can be computed proceeding as in \cite{Janickia,Janickib}. More
exactly, we get 
\begin{subequations}
\label{SOL}
\begin{align}
\frac{3}{2}\gamma B_{0}B^{2}+\gamma B_{0}^{3}+\frac{3}{2}\gamma
B_{0}C^{2}-F_{0}+\frac{3}{4}\gamma B^{2}C\cos 2\varphi & =0,  \label{f1} \\
\zeta B\Omega -3\gamma B_{0}BC\sin 2\varphi & =0,  \label{f2} \\
\frac{1}{4}B\Omega ^{2}-\frac{3}{4}\gamma B^{3}-3\gamma B_{0}^{2}B-\frac{3}{2%
}\gamma BC^{2}-3\gamma B_{0}BC\cos 2\varphi & =0,  \label{f3}
\end{align}%
where 
\begin{equation}
C=\dfrac{F}{\frac{3}{4}\gamma A^{2}-\Omega ^{2}}.  \label{C}
\end{equation}
\end{subequations}
We note that in papers \cite{Kovacic2011,Szemplinska1986} a form describing
a combination of $1:1$ and $1:2$ resonances was assumed (see Eq. (8.5.20) in 
\cite{Kovacic2011} and Eq. (8a) in \cite{Szemplinska1986}) and thus
different equations for the asymptotic solution were obtained.

Assuming $B\neq 0$ we get from Eqs. (\ref{f2}), (\ref{f3})
\begin{equation}
\left. 
\begin{array}{lll}
S_{1}\left( B_{0},B,\Omega ;\zeta ,\gamma ,F\right) & = & \zeta ^{2}\Omega
^{2}+\left( \frac{1}{4}\Omega ^{2}-\frac{3}{4}\gamma B^{2}-3\gamma B_{0}^{2}-%
\frac{3}{2}\gamma C^{2}\right) ^{2} \\ 
&  & -9\gamma ^{2}B_{0}^{2}C^{2}=0,%
\end{array}%
\right.  \label{S_1}
\end{equation}

Moreover, equations (\ref{f1}) and  (\ref{f3}) lead to
\begin{equation}
\left. 
\begin{array}{lll}
S_{2}\left( B_{0},B,\Omega ;\zeta ,\gamma ,F_{0},F\right) & = & -B^{2}\Omega
^{2}+3B^{4}\gamma -12B^{2}\gamma B_{0}^{2}+6B^{2}\gamma C \\ 
&  & -16\gamma B_{0}^{4}-24\gamma B_{0}^{2}C^{2}+16B_{0}F_{0}=0.%
\end{array}%
\right.  \label{S_2}
\end{equation}

Equation (\ref{S_1}) is quadratic concerning $B_{0}^2$. Therefore, we
solve this equation for $B_{0}^{2}$
\begin{equation}
\left. 
\begin{array}{l}
B_{0}^{2}=-\frac{1}{4}B^{2}+\frac{1}{12\gamma }\Omega ^{2}\pm \dfrac{\sqrt{%
f\left( B,\Omega ;\zeta ,\gamma ,F\right) }}{3\gamma \left( 4\Omega
^{2}-3\gamma B^{2}\right) ^{2}} \\ 
f\left( B,\Omega ;\zeta ,\gamma ,F\right) =-81\Omega ^{2}\zeta ^{2}\gamma
^{4}B^{8}+108\gamma ^{3}\left( 4\Omega ^{4}\zeta ^{2}-3F^{2}\gamma \right)
B^{6} \\ 
-108\Omega ^{2}\gamma ^{2}\left( 8\Omega ^{4}\zeta ^{2}-9F^{2}\gamma \right)
B^{4}+96\gamma \Omega ^{4}\left( 8\Omega ^{4}\zeta ^{2}-9F^{2}\gamma \right)
B^{2} \\ 
-256\zeta ^{2}\Omega ^{10}+192F^{2}\gamma \Omega ^{6}-576F^{4}\gamma ^{2}%
\end{array}%
\right.  \label{substitution}
\end{equation}

We substitute the following expression for $B_{0}$
\begin{equation}
B_{0}\left( B,\Omega \right) =\sqrt{-\frac{1}{4}B^{2}+\frac{1}{12\gamma }%
\Omega ^{2}-\frac{\sqrt{f\left( B,\Omega ;\zeta ,\gamma ,F\right) }}{3\gamma
\left( 4\Omega ^{2}-3\gamma B^{2}\right) ^{2}}}  \label{A_0}
\end{equation}%
to Eq. (\ref{S_2}) (it turns out that we have to choose the minus sign) to
get a complicated but useful implicit non-polynomial function $L\left( B,\Omega ;\zeta
,\gamma ,F_{0},F\right) =0$
\begin{equation}
L\left( B,\Omega ;\zeta ,\gamma ,F_{0},F\right) =S_{2}\left( B_{0}\left(
B,\Omega \right) ,B,\Omega ;\zeta ,\gamma ,F_{0},F\right) .  \label{L}
\end{equation}

\section{Vertical tangencies and singular points}
\label{singular}

Equations for vertical tangencies read
\begin{subequations}
\label{VT}
\begin{eqnarray}
L\left( B,\Omega ;\zeta ,\gamma ,F_{0},F\right) &=&0,  \label{vt1a} \\
\frac{\partial L\left( B,\Omega ;\zeta ,\gamma ,F_{0},F\right) }{\partial B}
&=&0,  \label{vt1b}
\end{eqnarray}
\end{subequations}

while  equations for singular points are
\begin{subequations}
\label{GenSing}
\begin{eqnarray}
L\left( B,\Omega ;\zeta ,\gamma ,F_{0},F\right) &=&0,  \label{gs1a} \\
\frac{\partial L\left( B,\Omega ;\zeta ,\gamma ,F_{0},F\right) }{\partial B}
&=&0,  \label{gs1b} \\
\frac{\partial L\left( B,\Omega ;\zeta ,\gamma ,F_{0},F\right) }{\partial
\Omega } &=&0.  \label{gs1c}
\end{eqnarray}%
\end{subequations}
Equations (\ref{VT}), (\ref{GenSing} can be solved numerically, yet simplify greatly for $B=0$.

We check that $\left[ \partial L\left( B,\Omega ;\zeta ,\gamma
,F_{0},F\right) /\partial B\right] _{B=0}$.
Therefore, we obtain a simplified equation for vertical tangencies
\begin{equation}
L\left( 0,\Omega ;\zeta ,\gamma ,F_{0},F\right) =0.  \label{vt2}
\end{equation}
\newpage
Equation (\ref{vt2}) can be solved for $\Omega $ yielding 
\begin{subequations}
\label{F1F2}
\begin{eqnarray}
f_{1}\left( \Omega ;\zeta ,\gamma ,F_{0},F\right)  &=&\Omega ^{12}+16\lambda
^{2}\Omega ^{10}-12\gamma F^{2}\Omega ^{6}+36\gamma ^{2}F^{4}=0,  \label{f_1} \\
f_{2}\left( \Omega ;\zeta ,\gamma ,F_{0},F\right) 
&=&\sum\limits_{k=0}^{18}a_{k}\Omega ^{2k}=0,  \label{f_2}
\end{eqnarray}
\end{subequations}
where non-zero coefficients $a_k$ are shown in Table \ref{tab:T1}.
\begin{table}[h!]
\caption{Non-zero coefficients $a_{k}$ of the polynomial (\ref{f_2})}
\label{tab:T1}\centering
\begin{tabular}{|l|l|}
\hline
$k$ & $a_{k}$ \\ \hline \hline
$18$ & $1$ \\ \hline
$17$ & $48\zeta ^{2}$ \\ \hline
$16$ & $768\zeta ^{4}$ \\ \hline
$15$ & $36\gamma F^{2}-3456\gamma F_{0}^{2}+4096\zeta ^{6}$ \\  \hline
$14$ & $1152\zeta ^{2}\gamma F^{2}+165\,888\zeta ^{2}\gamma F_{0}^{2}$ \\ \hline
$13$ & $9216\gamma F^{2}\zeta ^{4}$ \\ \hline
$12$ & $756\gamma ^{2}F^{4}-248\,832\gamma
^{2}F^{2}F_{0}^{2}+2985\,984\gamma ^{2}F_{0}^{4}$ \\ \hline
$11$ & $24\,192\zeta ^{2}\gamma ^{2}F^{4}+1990\,656\gamma
^{2}F^{2}F_{0}^{2}\zeta ^{2}$ \\ \hline
$10$ & $193\,536\gamma ^{2}F^{4}\zeta ^{4}$ \\ \hline
$9$ & $3456\gamma ^{3}F^{6}-2239\,488\gamma ^{3}F^{4}F_{0}^{2}$ \\ \hline
$8$ & $165\,888\gamma ^{3}F^{6}\zeta ^{2}$ \\ \hline
$6$ & $-31\,104\gamma ^{4}F^{8}+4478\,976\gamma ^{4}F_{0}^{2}F^{6}$ \\ \hline
$5$ & $2488\,320\gamma ^{4}F^{8}\zeta ^{2}$ \\ \hline
$3$ & $-933\,120F^{10}\gamma ^{5}$ \\ \hline
$0$ & $4665\,600F^{12}\gamma ^{6}$ \\ \hline
\end{tabular}
\end{table}

Moreover, equation for singular points (\ref{GenSing}) in the case $B=0$ can
be significantly simplified, demanding that the equation (\ref{f_2}) has a double
root (equation (\ref{f_1}) has no physical double roots). We can request
that the discriminant of the polynomial $f_{2}\left( \Omega \right) $
vanishes or solve the equivalent set of equations
\begin{subequations}
\label{SING}
\begin{eqnarray}
f_{2}\left( \Omega ;\zeta ,\gamma ,F_{0},F\right)  &=&0,  \label{s1a} \\
\frac{\partial f_{2}\left( \Omega ;\zeta ,\gamma ,F_{0},F\right) }{\partial
\Omega } &=&0.  \label{s1b}
\end{eqnarray}
\end{subequations}
By solving Eqs. (\ref{SING}) for $F_{0},\ F$ we obtain clear and simplified equations for
singular points, making the computations easier, 
\begin{equation}
p\left( Z\right)
=d_{12}Z^{12}+d_{10}Z^{10}+d_{8}Z^{8}+d_{6}Z^{6}++d_{4}Z^{4}+d_{2}Z^{2}+d_{0}=0,
\label{p(Z)}
\end{equation}

\vspace{-0.5cm}

\begin{center}
$%
\begin{array}{l}
d_{12}=466\,560\gamma ^{6},\ d_{10}=-233\,280\gamma ^{5}\Omega ^{2},\
d_{8}=20\,736\gamma ^{4}\Omega ^{2}\left( 2\Omega ^{2}+11\zeta ^{2}\right) ,
\\ 
d_{6}=-1728\gamma ^{3}\Omega ^{4}\left( 2\Omega ^{2}+41\zeta ^{2}\right) ,\
d_{4}=54\gamma ^{2}\Omega ^{4}\left( 3\Omega ^{4}+112\zeta ^{2}\Omega
^{2}+448\zeta ^{4}\right) , \\ 
d_{2}=-3\gamma \Omega ^{6}\Omega ^{4}+72\zeta ^{2}\Omega ^{2}+896\zeta
^{4},\ d_{0}=4\zeta ^{2}\Omega ^{6}\left( \Omega ^{4}+20\zeta ^{2}\Omega
^{2}+64\zeta ^{4}\right) .%
\end{array}%
$
\end{center}

where $Z=\dfrac{F}{\Omega ^{2}}$, and 

\begin{equation}
q\left( T\right) =e_{2}T^{2}+e_{0}=0,  \label{q(T)}
\end{equation}

\vspace{-0.5cm}

\[
\left. 
\begin{array}{l}
e_{2}=1728\gamma \Omega ^{6}+41\,472\gamma \zeta ^{2}\Omega
^{4}+1603\,584\gamma \zeta ^{4}\Omega ^{2}+22\,118\,400\gamma \zeta
^{6},\medskip  \\ 
e_{0}=-\Omega ^{14}+\left( -534Z^{2}\gamma +592\zeta ^{2}\right) \Omega
^{12}+\left( 
\begin{array}{l}
15\,768Z^{4}\gamma ^{2}-31\,968Z^{2}\zeta ^{2}\gamma  \\ 
+20\,128\zeta ^{4}%
\end{array}%
\right) \Omega ^{10} \\ 
+\left( -247\,968Z^{6}\gamma ^{3}+627\,552Z^{4}\zeta ^{2}\gamma
^{2}-499\,008Z^{2}\zeta ^{4}\gamma +195\,072\zeta ^{6}\right) \Omega ^{8} \\ 
+\left( 
\begin{array}{l}
1736\,640Z^{8}\gamma ^{4}-3805\,056Z^{6}\zeta ^{2}\gamma
^{3}+2596\,608Z^{4}\zeta ^{4}\gamma ^{2} \\ 
-958\,464Z^{2}\zeta ^{6}\gamma +458\,752\zeta ^{8}%
\end{array}%
\right) \Omega ^{6} \\ 
+\left( 
\begin{array}{l}
-4199\,040Z^{10}\gamma ^{5}+9642\,240Z^{8}\zeta ^{2}\gamma
^{4}+7617\,024Z^{6}\zeta ^{4}\gamma ^{3} \\ 
-40\,255\,488Z^{4}\zeta ^{6}\gamma ^{2}+16\,465\,920Z^{2}\zeta ^{8}\gamma 
\end{array}%
\right) \Omega ^{4} \\ 
+\left( -19\,906\,560Z^{10}\zeta ^{2}\gamma ^{5}-76\,308\,480Z^{8}\zeta
^{4}\gamma ^{4}+23\,224\,320Z^{6}\zeta ^{6}\gamma ^{3}\right) \Omega ^{2} \\ 
-99\,532\,800Z^{10}\zeta ^{4}\gamma ^{5}.%
\end{array}%
\right. 
\]
where $T=F_{0}\Omega $.

\section{Numerical verification}
\label{numerical}
In this Section, we shall compute singular points as well as vertical
tangencies for chosen values of $\zeta $, $\gamma $, and $\Omega $. In the
case $B=0$, we use Eqs. (\ref{SING}) in reduced form (\ref{p(Z)}) and (\ref{q(T)})
 to compute singular points and Eqs. (\ref{F1F2}) to compute vertical
tangencies. Then, in the case $B\neq 0$, we use Eqs. (\ref{GenSing}) and (\ref%
{VT}), respectively.

In what follows we choose for $B = 0$, quite arbitrarily, $\zeta =0.09$, $\gamma =0.3$, $%
\Omega =1.5$.

\subsection{Singular points and vertical tangencies, $B=0$}
\label{B=0}
Thus, the singular point is chosen as $\left( B,\,\Omega \right)
=\left( 0,\,1.5\right) $. We need to compute the parameters, $F_{0}$ and $F$,
for which the selected point is singular.

Therefore, for $\zeta =0.09$, $\gamma =0.3$, $\Omega =1.5$, and $B=0$, we
solve equation $p\left( Z\right) =0$, with $p\left( Z\right) $ defined in
Eq. (\ref{p(Z)}), obtaining four positive roots, $Z=0.\,198\,445$, $%
0.\,509\,084$, $1.\,095\,980$, $1.\,259\,811$. We check, however, that only $%
Z=1.\,095\,980$ leads to a solution of (\ref{GenSing}). Since $Z=F/\Omega
^{2}$, we get, for $Z=1.\,095\,980$, $F=2.465\,955$. We now solve $q\left(
T\right) =0$, with $q\left( T\right) $ defined in Eq. (\ref{q(T)}),
obtaining for $Z=1.\,095\,980$ one positive root $T=0.112\,203$. Since $%
T=F_{0}\Omega $ we get $F_{0}=0.074\,802\,$.

Now, we check that for $\gamma =0.3$, $F_{0}=0.074\,802$, $F=2.465\,955$
equations (\ref{GenSing}), solved numerically, yield indeed $\zeta =0.09$
and an isolated singular point $\left( B,\,\Omega \right) =\left(
0,\,1.5\right) $, see Fig. \ref{F1}.

To find vertical tangencies, we set, for example, $\zeta =0.082$ and use just
computed parameter values: $\gamma =0.3$, $F_{0}=0.074\,802$, $F=2.465\,955$%
. Solving equation (\ref{f_2}), we get $\left( B,\,\Omega \right) =\left(
0,\,1.\,474\,612\right) $, $\left( 0,\,1.\,527\,914\right) $ (alll solutions
of Eq. (\ref{f_1} are complex); see red boxes in Fig. \ref{F1}.

\newpage

\subsection{Singular points and vertical tangencies, $B\neq 0$}
\label{B>0}

We work with parameter values computed in the preceding subsection, i.e., $%
\gamma =0.3$, $F_{0}=0.074\,802$, $F=2.465\,955$. We have shown in the
prior Section that equations (\ref{GenSing}) have for $\zeta =0.09$ an
isolated singular point $\left( B,\Omega \right) =\left( 0,1.5\right) $.

Moreover, equations (\ref{GenSing}) have other singular points for $\gamma
=0.3$, $F_{0}=0.074\,802$, $F=2.465\,955$, and $B\neq 0$. Solving
numerically Eqs. (\ref{GenSing}) we get (i) $\zeta =0.086\,504$ and a pair
of self-intersections $\left( B,\Omega \right) =\left( \pm
0.305\,755,1.\,496\,498\right) $, (ii) a pair of isolated points, $\zeta
=0.108\,010$, $\left( B,\ \Omega \right) =\left( \pm 1.\,069\,257,\
1.\,\allowbreak 613\,277\right) $ , see Fig. \ref{F1} where we show all singular
points. 

We also compute vertical tangencies for $B\neq 0$. Solving equations (\ref%
{vt1a}) and (\ref{vt1b}) numerically for $\gamma $, $F_{0}$, $F$ listed above
and $\zeta =0.082$ we get $\left( B,\Omega \right) =\left( \pm
0.318\,833,1.\,514\,052\,\right) $; see red boxes in Fig. \ref{F1}.

\subsection{Amplitude-frequency plots and bifurcation diagrams}
\label{plots}
We can show all singular points computed for $1:2$ resonance  in one plot.

\begin{figure}[h!]
\center
\includegraphics[width=11cm, height=7cm]{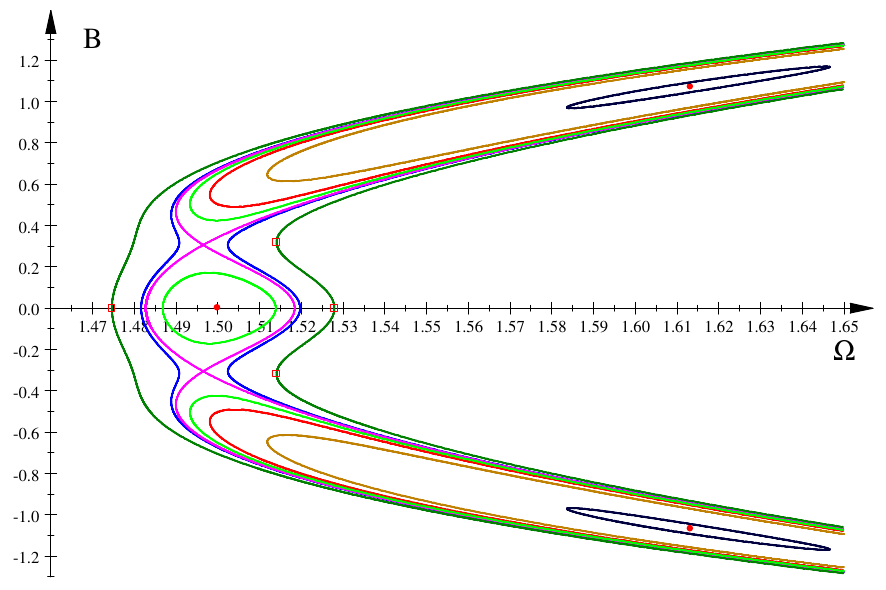}
\caption{Sequential metamorphoses of amplitude-frequency implicit function $L\left( B,\Omega ;\zeta ,\gamma ,F_{0},F\right) = 0$, describing $1:2$ resonance.
}
\label{F1}
\end{figure}

Parameters in Fig. \ref{F1} are $\gamma =0.3$, $F_{0}=0.07480195$, $F=2.\,465\,954$, and $%
\zeta =0.108\,010$ (two red dots), $0.107\,2$ (navy), $0.095$ (sienna), $
0.09$ (red, a dot and two branches), $0.088$ (light green, an oval and two branches), $
0.086\,504$ (magenta, two self-intersections), $0.086$ (blue), $0.082$
(green). 

We have computed bifurcation diagrams solving Eq. (\ref{Asym-Duffing})
numerically -- obtaining $y\left( t\right) $ as a function of $\Omega $; see
figures below. Comparison with Fig. \ref{F1} reveals which branches are
stable. Colors in bifurcation diagrams correspond to those in Fig. \ref{F1}.
We thus document metamorphoses of $1:2$ resonance (two branches, colored) and its
interaction with the primary $1:1$ resonance (one branch in the middle, black).

\begin{figure}[h!]
\center
\includegraphics[width=12cm, height=8cm]{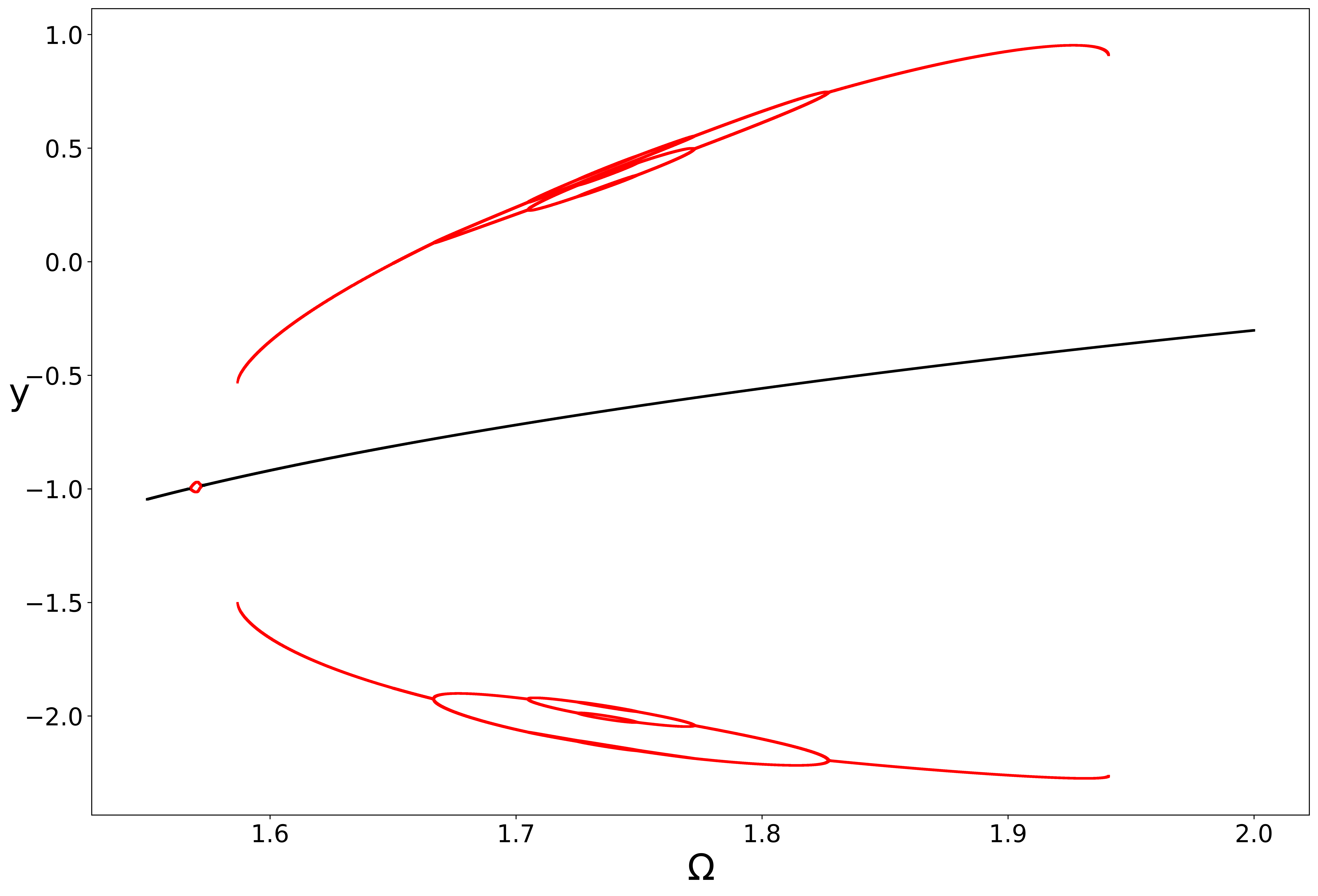}
\includegraphics[width=12cm, height=8cm]{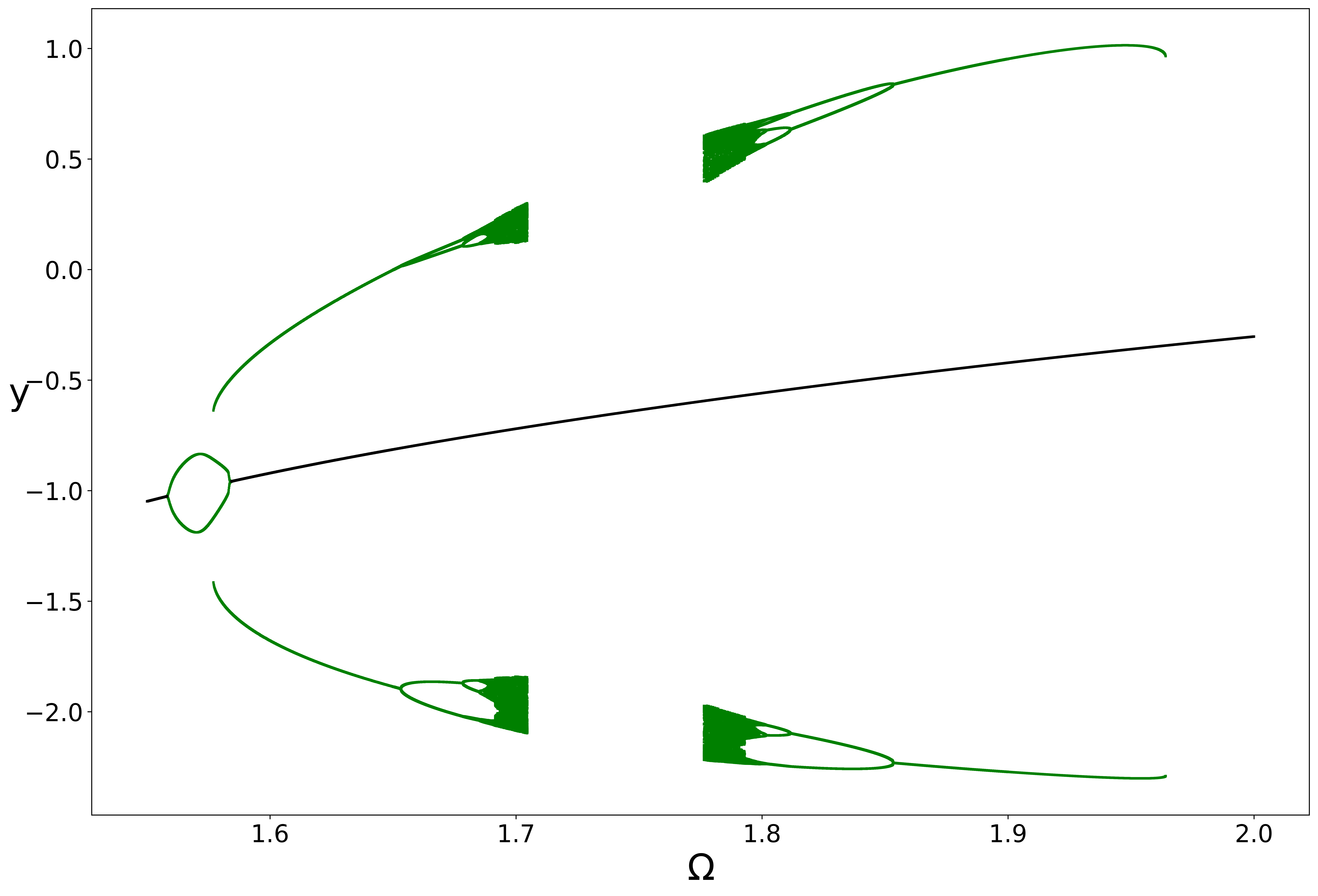}
\caption{Bifurcation diagrams, 
$\zeta = \,0.094\, 185$ (top), 
$\zeta =0.091$ (bottom).
}
\label{F2}
\end{figure}

\begin{figure}[h!]
\center

\includegraphics[width=12cm, height=8cm]{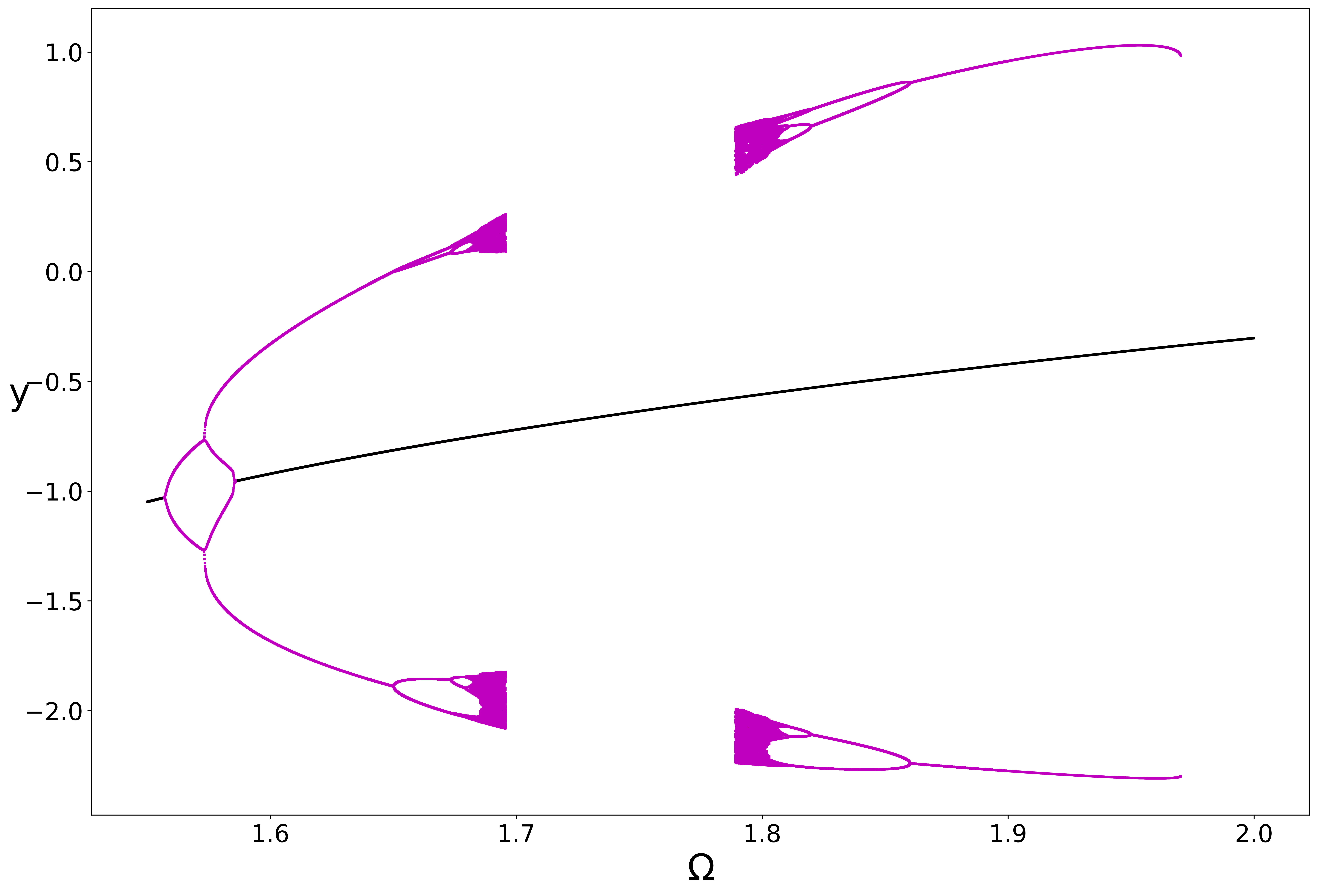}
\includegraphics[width=12cm, height=8cm]{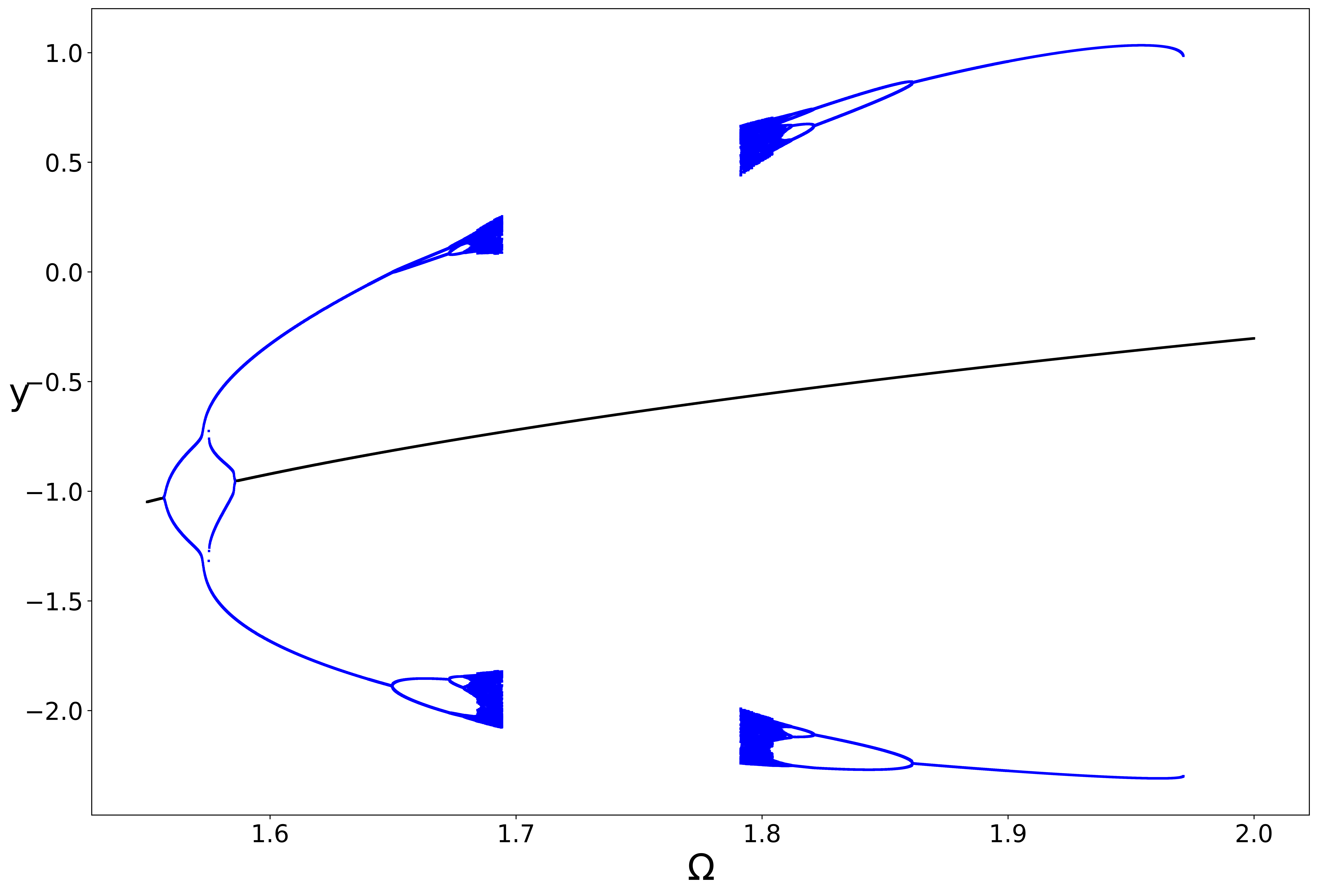}
\caption{Bifurcation diagrams, $\zeta = 0.090\, 14$ (top), 
 $\zeta = 0.90$ (bottom).
}
\label{F3}
\end{figure}

\subsection{Description of transmutations of $1:2$ resonance}
\label{description}

We describe metamorphoses of $1:2$ resonance for  $\gamma =0.3$, $F_{0}=0.074\,802$, $F=2.465\,955$ 
and descending values of $\zeta$. In what follows, $\zeta_n$ denotes values of $\zeta$ computed integrating 
numerically the Duffing equation (\ref{Asym-Duffing}) while $\zeta_a$ means values of $\zeta$ computed 
from analytical condition (\ref{GenSing}) for singular points. We have computed function 
$L\left( B,\Omega ;\zeta ,\gamma ,F_{0},F\right)$ from the asymptotic solution (\ref{Solution 1/2}) and (\ref{SOL}).

In Fig. \ref{F1}, transmutations of  function  $L\left( B,\Omega ;\zeta ,\gamma ,F_{0},F\right)$ are shown, 
and in bifurcation diagrams \ref{F2} and \ref{F3}, metamorphoses of solutions of Duffing equation 
(\ref{Asym-Duffing}), are presented.  We note good qualitative correspondence between predicted 
transmutations shown in Fig. \ref{F1} and metamorphoses of the solutions of the Duffing equation 
documented in Figs. \ref{F2} and \ref{F3}.

\smallskip
We describe these changes as follows.

\begin{enumerate}
\item For $\zeta _{n}=0.114\,69$ ($\zeta _{a}=0.108\,010$), $1:2$ resonance
appears for the first time at $\Omega_n = 1.722$ ($\Omega_a = 1.613$); 
see two red dots (singular isolated points) in Fig. \ref{F1} (we do not show the corresponding 
bifurcation diagrams).

For descending values of $\zeta$, the $1:2$ resonance grows and transforms rapidly. 
More precisely, two inverted cascades form, then break and pull out -- we discuss 
these changes in Section \ref{test}.

\item For $\zeta _{n}=0.0941\,85$  ($\zeta
_{a}=0.09$), a new isolated point of $1:2$ resonance (red) appears on $1:1$
resonance (black) at $\Omega_n = 1.57$ ($\Omega_a =1.5 $); see  small red circle in top figure \ref{F2} or 
red dot in Fig. \ref{F1}. The primary $1:1$ resonance is black.

It is the first contact of these resonances. 
[Note that the 1 2 resonance has been subject to two-period doublings but has not broken yet.]

\item The singular, isolated point gives rise to an oval -- a period- doubling 
of $1:1$ resonance; see the light green line in Fig. \ref{F1} and green oval in bottom figure \ref{F2}.
The $1:1$ resonance is black. 

There is still no contact between the primer $1:2$ resonance and the primary resonance 
[there are already two whole cascades of period-doublings of $1:2$ resonance that have been 
disrupted and moved away; we call them left and right; see green lines in Fig. \ref{F2})].

\item For $\zeta _{n}=0.090\,14$ ($\zeta _{a}=0.086\,504$), there are two
self-intersections (two singular points) at $\Omega_n = 1.573$ ($\Omega_a = 1.496$)
and resonance $1:1$ and left resonance $1:2$ merge, see magenta lines in Fig. \ref{F1} and top figure \ref{F3}.
The right $1:2$ resonance stays separated. The primary resonance is black. 

\item For $\zeta <\zeta _{n}=0.090\,14$ ($\zeta _{a}<0.086\,504$), connected
branches split another way; see blue lines in Fig. \ref{F1} and Fig. \ref{F3}.  
The primary $1:1$ resonance (black) absorbs the left $1:2$ resonance with the whole cascade of period doublings, 
there is also another branch of $1:1$ resonance, with one period doubling. The right $1:2$ resonance stays unchanged. 

\end{enumerate}

Summing up, the separate $1:2$ resonance, after the first contact with $1:1$ resonance, splits into two parts, 
left and right, then the left resonance $1:2$ merges with the primary $1:1$ resonance and splits another way, 
while the right $1:2$ resonance does not evolve.

\section{On the formation of period-doubling cascades and their subsequent breaking}
\label{test}

In the preceding section, we described complicated interactions of $1:2$ and $1:1$ resonances, which are generic 
for dynamical system \ref{Asym-Duffing}. We thus decided to investigate the generality of our scenario by computing 
bifurcation diagrams for another resonance in another dynamical system.

More precisely, we analyzed a vibrating system with two minima of 
potential energy, studied by Szempli\'{n}ska-Stupnicka
\begin{equation}
\frac{d^{2}x}{dt^{2}}+h\frac{dx}{dt}-\frac{1}{2}x+\frac{1}{2}x^{3}=F\cos
\left( \omega t\right),   \label{Sz}
\end{equation}
see Eq. (4.2) in  \cite{Szemplinska2003}.
\begin{figure}[h!]
\center
\includegraphics[width=12cm, height=5.5cm]{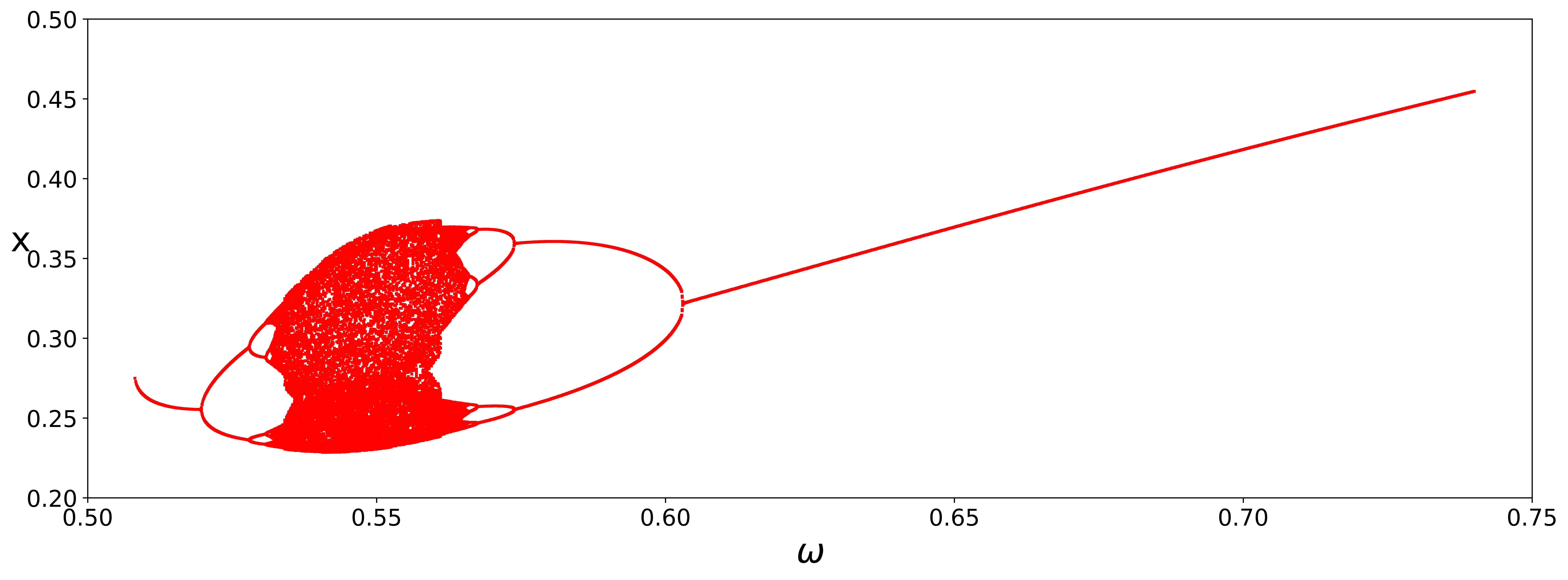}
\includegraphics[width=12cm, height=5.5cm]{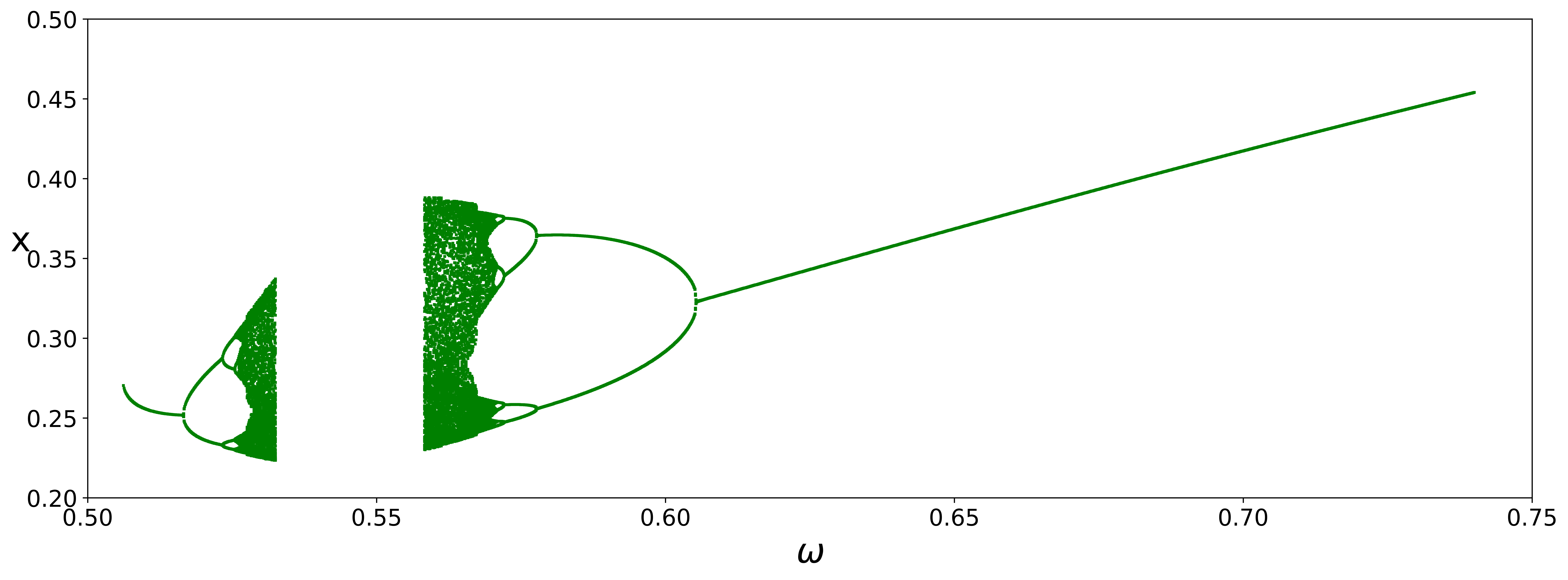}
\caption{Bifurcation diagrams for Eq. (\ref{Sz}), $F = 0.0620,  h=0.1259$ (red, top), 
 $F = 0.0620,  h=0.1258$  (green, bottom).
}
\label{F4}
\end{figure}
We computed bifurcation diagrams for the $1:1$ resonance of Eq. (\ref{Sz}) 
for F = 0.0620, h = 0.1259 (red), and F = 0.0620,h = 0.1259 (green). We see that fully developed cascade breaks in 
close analogy to transformation shown in Fig. \ref{F2} for the $1:2$ resonance. We attribute disruption of the cascade 
to crisis, cf. comments below Fig. 4.15  in \cite{Szemplinska2003} (it seems that for $h=0.1$ the second cascade 
was destroyed completely). 

It seems typical to break a fully developed cascade by forming two cascades. Note that in this case, one of these cascades 
may remain unnoticed.

\section{Summary and most important findings}
\label{summary}
Based on the amplitude-frequency steady-state implicit equation (\ref{L}) computed 
for Eq. (\ref{Asym-Duffing}), we studied the metamorphoses of the resonance $1:2$ and its 
interaction with the primary resonance  $1:1$. 

Working in the formalism of differential properties of implicit functions, we derived formulas to 
compute singular points of the amplitude-frequency function $L\left( B,\Omega ;\zeta ,\gamma ,F_{0},F\right) $, 
defined in (\ref{L}); see Section \ref{singular}. 
It should be stressed that the dynamics of the initial equation (\ref%
{Asym-Duffing}) change in the neighborhood of singular points (in the
parameter space). 

In Section \ref{numerical}, after choosing arbitrary parameters $\zeta =0.09$, $\gamma =0.3$, 
$\Omega =1.5$, and $B=0$, we (i) first solved Eqs. (\ref{SING}), (\ref{q(T)}) obtaining
$F=2.465\,955$, $F_{0}=0.074\,802$. We thus (ii) computed parameters of the singular point 
$\left( B,\ \Omega \right) =\left( 0,\ 1.5\right) $; see red dot in Fig. \ref{F1}. Other singular points 
were computed numerically from Eqs. (\ref{GenSing}). See two magenta  self-intersections and two 
red dots in  Fig. \ref{F1}. 

Working in this order is relatively easy as we know the position of the first singular point 
$\left( B,\ \Omega \right) =\left( 0,\ 1.5\right) $; thus two other pairs of singular points 
should not be too far away on the $\Omega$ axis. This information is of great help 
in solving  Eqs. (\ref{GenSing}) numerically.

We also computed bifurcation diagrams solving Eq. (\ref{Asym-Duffing}) numerically, shown in Figs. 
\ref{F2} and \ref{F3}, obtaining good agreement with the amplitude-frequency profiles of Fig. \ref{F1}. 
Subsection \ref{description} provides a detailed description of metamorphoses of $1: 2$ resonance.
Section \ref{test} shows another example of cascade disruption. 

The most significant achievements of this work, consisting of computing singular points of the amplitude-frequency 
implicit function (\ref{L}), are 

\begin{enumerate}

\item The semi-analytic procedure to compute singular, isolated points of $1:2$ resonance, 
consisting of Eqs. (\ref{p(Z)}) and (\ref{q(T)}). These
points correspond to the first contact of resonances $1:2$ and $1:1$.
\newline Numerical computation of a pair of isolated points that indicate the birth
of $1:2$ resonance. 

\item Discovery and detailed description of complicated transmutations of $1:2$ resonance that resulted in 
the first contact with the primary $1:1$ resonance, disruption of the $1:2$ resonance into two parts, 
left and right, merging of the left $1:2$ resonance with the primary $1:1$ resonance, and breaking again; 
the right $1:2$ resonance effectively not evolving.

\item Discovery of analogous metamorphoses in a different dynamical system, suggesting a greater generality 
of our results; see Section \ref{test}.

\end{enumerate}

\appendix
\label{detals}
\section{Computational details}

Nonlinear equations were solved numerically using the
computational engine Maple 4.0 from Scientific WorkPlace 4.0.

 Figure \ref{F1} was plotted with the computational engine MuPAD 4.0 from Scientific WorkPlace 
5.5. Bifurcation diagrams in Figs. \ref{F2}, \ref{F3}, and \ref{F4}  were computed by integrating 
numerically Eq.~(\ref{Asym-Duffing}) running DYNAMICS \cite{Nusse2012} as well as 
our programs written in Pascal and Python \cite{Perez2007}.

\end{document}